# VAEMax: Open-Set Intrusion Detection based on OpenMax and Variational Autoencoder

Zhiyin Qiu[1,2], Ding Zhou[1], Yahui Zhai[3], Bo Liu[1,2], Lei He[1], Jiuxin Cao[1,2]*
[1]*Purple Mountain Laboratories*, Nanjing, China
[2]*School of Cyber Science and Engineering, Southeast University*, Nanjing, China
[3]*TravelSky Technology Limited*, Beijing, China
*Corresponding author's email: jx.cao@seu.edu.cn

*Abstract*—Promptly discovering unknown network attacks is critical for reducing the risk of major loss imposed on system or equipment. This paper aims to develop an open-set intrusion detection model to classify known attacks as well as inferring unknown ones. To achieve this, we employ OpenMax and variational autoencoder to propose a dual detection model, VAEMax. First, we extract flow payload feature based on one-dimensional convolutional neural network. Then, the OpenMax is used to classify flows, during which some unknown attacks can be detected, while the rest are misclassified into a certain class of known flows. Finally, use VAE to perform secondary detection on each class of flows, and determine whether the flow is an unknown attack based on the reconstruction loss. Experiments performed on dataset CIC-IDS2017 and CSE-CIC-IDS2018 show our approach is better than baseline models and can be effectively applied to realistic network environments.

## I. INTRODUCTION

New cybersecurity attacks are continuously spawned along with the fast evolution of network technology and network services. Therefore, it is becoming increasingly important to develop advanced intrusion detection systems to deal with the challenges brought by the rapid growth of malicious network attacks, especially unknown new attacks. However, traditional intrusion detection studies, such as misuse detection, are based on the closed assumption that all malicious and benign behaviors in the test set have occurred during the training phase. Anomaly detection relies on the semi-closed assumption that new attacks will occur during the testing phase, but no new benign behavior will emerge. Unfortunately, the real cyber scene is open, not closed, with new applications, services, and attacks emerging every day. This paper aims to develop an open-set network attack detection model capable of inferring unknown network attacks while performing fine-grained known network intrusion detection.

To the best our knowledge, most of the current open-set network attack detection methods are based on OpenMax or autoencoder. Zhang *et al.* [1] proposed the Open-CNN based on OpenMax. Yang *et al.* [2] proposed an unknown attack detection method combining conditional variational autoencoder (CVAE) with extreme value theory (EVT). Ping *et al.* [3] proposed the OpenIDS, including a MinMax autoencoder, classifier, and pseudo extreme value machine. Fu *et al.* [4] proposed an approach with an attention encoding and deep metric learning model for intrusion detection. Soltani *et al.* [5] identified unknown samples based on the Deep Open Classification [6]. However, these works are only based on one of the methods and do not combine the two methods. This prompted us to build a dual detection model that combines OpenMax and variational autoencoder (VAE). Our basic idea is to first classify flows based on OpenMax, during which some unknown attacks can be detected, while the rest are misclassified into a certain class of known flows. Then, use VAE to perform secondary detection on each class of flows, and determine whether the flow is an unknown attack based on the reconstruction loss. The main contributions of this paper are summarized as follows:

1) We propose the dual intrusion detection model, VAEMax, based on OpenMax and VAE. First, use OpenMax to classify flows. Then, use VAE to perform further detection on each class of flows, and determine whether the flow is an unknown attack based on the reconstruction loss.

2) We proposed a lightweight flow payload feature extraction method based on one-dimensional convolutional neural network (1D-CNN), and designed five auxiliary information to enhance the representation ability of payload information.

3) Experiments are conducted on two widely used datasets to evaluate open-set intrusion detection. The experimental results demonstrate that the proposed dual detection structure can achieve improved performance.

The remainder of the paper is organized as follows. Section II discusses related works. Section III employs OpenMax and VAE for building dual intrusion detection model. Section IV presents experimental results to show the performance of our proposed solution. Section V draws conclusions.

## II. RELATED WORK

### A. Payload Feature Extraction

With the rapid development of deep learning, researchers begin to apply convolution neural network (CNN), long short-term memory networks (LSTM), Transformer, autoencoder (AE) and other models to network flow payload feature extraction [7]. Aceto *et al.* [8,17] exploited the heterogeneity of traffic data by learning intra-modal and inter-modal dependencies, which overcame performance limitations of existing (myopic) single-modal deep learning-based traffic classification proposals. Liu *et al.* [9] proposed the FS-Net which adopts a multi-layer encoder-decoder structure which can mine the potential sequential characteristics of flows deeply, and import the reconstruction mechanism which can enhance the effectiveness of features. Lotfollahi *et al.* [10] employs stacked AE and CNN in order to classify network traffic. Rago *et al.* [11] adopted AE as key building blocks of the proposed Multi-Task Learning method for common feature representations. Liu *et al.* [12] proposed the BGRUA which uses bidirectional gate recurrent unit to extract the forward and backward features of the byte sequences in a session and adopts attention mechanism to assign weights to features according to their contributions to classification. Cheng *et al.* [13] proposed a lightweight model which adopts the multi-head attention and the 1D CNN. He *et al.* [14] proposed the PERT to perform automatic traffic feature extraction using a state-of-the-art dynamic word embedding technique, and utilized unlabeled traffic to pre-train an encoding network that learns the contextual distribution of traffic payload bytes. Xie *et al.* [15] proposed a self-attentive method for traffic classification where self-attention mechanism was utilized for interpretability exploration. Ren *et al.* [16] proposed a tree structural recurrent neural network, which divides a large classification into small classifications by using the tree structure. Lin *et al.* [18] proposed the TSCRNN, which extracts abstract spatial features by CNN and then introduces stack bidirectional LSTM to learn the temporal characteristics. Meng *et al.* [19] preserved both semantic and byte patterns of each packet, and utilized contrastive loss with a sample selector to optimize the learned representations so that similar packets were closer in the latent semantic space. Yao *et al.* [20] modeled the time-series network traffic by the recurrent neural network and introduced the attention mechanism for assisting network traffic classification. Dai *et al.* [21] proposed the global-local attention data selection model which allows the model to efficiently extract features from multimodal input with a single-modal-like approach. Meng *et al.* [22] provided a generative pretrained model NetGPT which propose the multi-pattern network traffic modeling to construct unified text inputs and support both traffic understanding and generation tasks.

*B. Open-Set Recognition*

With the development of neural network research, network attack identification and classification tasks have shown excellent performance on various public data sets. However, in real scenarios, new network attacks often appear, and the characteristics of these attacks are often unknown. At this time, traditional classifiers will fail. The Open Set Recognition (OSR) task aims to solve this problem [23,24]. When the model is trained, the category information contained in the training set is incomplete. During testing, the model will encounter samples that do not belong to the training set categories. This requires the model to correctly classify samples of known classes and at the same time accurately identify samples of unknown classes that appear. Existing OSR models are based on modeling methods and can be roughly divided into two categories: discriminative models and generative models. Cevikalp *et al.* [25,26] added a constraint to the target class samples based on traditional SVM, and proposed the best-fitting hyperplane classifier (BFHC) model. The BFHC model directly forms a flat plate in the feature space, and the kernel method can be used to extend the BFHC model to nonlinear cases. Scheirer *et al.* [27] proposed a Compact Decaying Probability (CAP) model is, in which the probability of class membership decays as the point moves from known data to open space. At the same time, they also proposed a Weibull-calibrated SVM, which combines statistical extreme value theory (EVT) and uses two independent support vector machines to calibrate scores. Bendale *et al.* [28] proposed the OpenMax model as the first solution for open set deep networks. First, calculate the mean activation vector (MAV) of the penultimate layer of the network for each class, and then calculate the distance between the training sample and its corresponding class MAV, which is used to fit the independent Weibull distribution of each class, and finally redistribute the scores according to the Weibull distribution fitting activation vector value. Oza *et al.* [29] proposed an open-set recognition algorithm using class conditioned auto-encoders with novel training and testing methodologies. Encoder learns the first task following the closed-set classification training pipeline, whereas decoder learns the second task by reconstructing conditioned on class identity. Furthermore, reconstruction errors are modeled by EVT to find the threshold for identifying known/unknown class samples. Yoshihashi *et al.* [30] proposed Classification-Reconstruction learning for Open-Set Recognition (CROSR), which utilizes latent representations for reconstruction and enables robust unknown detection without harming the known-class classification accuracy. Ge *et al.* [31] proposed the generative OpenMax algorithm, using conditional generative adversarial network to synthesize mixed unknown samples. Neal *et al.* [32] reformulated the open set recognition problem as classification with one additional class, which includes the set of novel and unknown examples by generating examples that are close to training set examples yet do not belong to any training category. Perera *et al.* [33] trained a generative model for all known classes and then augmented the input with the representation obtained from the generative model to learn a classifier. Sun *et al.* [34] proposed the Conditional Gaussian Distribution Learning (CGDL) which forced different latent features to approximate different Gaussian models. Zhang *et al.* [35] proposed the OpenHybrid framework, including an encoder, a classifier, and a flow-based density estimator to detect whether a sample belongs to the unknown category. Zhou *et al.* [36] proposed to learn Placeholders for Open-Set Recognition (PROSER), which prepares for the unknown classes by allocating placeholders for both data and classifier. Kong *et al.* [37] proposed OpenGAN, which augmented the available set of real open training examples with adversarially synthesized "fake" data and built the discriminator over the features computed by the closed-world K-way networks. Lyu *et al.* [38] proposed MetaMax, a more effective post-processing technique that improved upon contemporary methods by directly modeling class activation vectors.

III. METHODOLOGY

In this section, we describe the VAEMax model. First, the overall architecture of VAEMax is described in subsection A. Then, we describe the implementation details and parameters setting of each component of VAEMax in subsection B to E.

*A. Overall architecture*

As mentioned in Section I, the core idea of the VAEMax model is dual detection. First, the OpenMax is used to classify flows, during which some unknown attacks can be detected, while the rest are misclassified into a certain class of known flows. Then, use VAE to perform secondary detection on each class of flows, and determine whether the flow is an unknown attack based on the reconstruction loss. The main components of the VAEMax are Payload Information Extraction, Payload Feature Encoder, OpenMax, Variational Autoencoder. Payload Feature Encoder consists of two layers of 1D-CNN. The VAEMax model training is divided into two stages, first train Payload Feature Encoder and OpenMax, then freeze the parameters of Payload Feature Encoder and train Variational Autoencoder.

*B. Payload Information Extraction*

Bidirectional flows are uniquely marked by five-tuple information <source IP, source port, destination IP, destination port, protocol>. For each flow, extract its first 16 packets, and each packet extracts its first 128 bytes, padding with zeros if insufficient. In addition, five auxiliary information are extracted for each packet, DIRC (flow direction), FLAG (TCP Flags), PLDL (payload length), TCPW (TCP window size), ITVT (packet interval time). Thus, the extracted payload information can be formalized as (1).

$$Flow = \{payload, auxiliary\},$$
$$payload = \{pkt_1, pkt_2, \cdots, pkt_{16}\},$$
$$pkt_i = \{x_1, x_2, \cdots, x_{128}\}, x_i \in [0,1],$$
$$auxiliary = \{aux_1, aux_2, \cdots, aux_{16}\}$$
$$aux_i = \{DIRC, FLAG, PLDL, TCPW, ITVT\}$$
$$FLAG = \{FIN, SYN, RST, PSH, ACK, URG, ECE, CWR\}$$
(1)

In order to obtain better performance, but also to facilitate neural network processing, all the packet bytes are divided by 255, the maximum value for a byte, so that all the input values are in the range [0, 1]. For the auxiliary information, we normalize it to fit the Standard Gaussian distribution, where the 8 flags in TCP Flags are processed separately. After the above preprocessing, the data is finally in the following form:

$$Flow = \left\{ \begin{pmatrix} p_{1,1} & \cdots & p_{1,128} \\ \vdots & \ddots & \vdots \\ p_{16,1} & \cdots & p_{16,128} \end{pmatrix}, \begin{pmatrix} a_{1,1} & \cdots & a_{1,12} \\ \vdots & \ddots & \vdots \\ a_{16,1} & \cdots & p_{16,12} \end{pmatrix} \right\} \quad (2)$$

*C. Payload Feature Encoder*

Different from the work in Section II, we use a very shallow encoder to extract payload features. The reason for this is that on the one hand, the real-time requirements in the actual network environment are taken into account, and on the other hand, the shallow network can already extract the deep features of the payload very well, which was confirmed in the experiment. Inspired by work on natural language processing, we treat the payload content in a packet as a text containing semantic information. Therefore, we use 1D-CNN to perform convolution operations on each packet to extract its hidden features. In order to speed up the convergence of the model and increase nonlinearity, we performed batch regularization and activation function operations after convolution. The specific model structure and parameters are shown in Payload Feature Encoder in Fig. 1. The four parameters of 1D-CNN are input channels, output channels, kernel size, stride. Thus, the payload feature extraction process can be formalized as follows:

$$PayloadFeature = CNNBlock2(CNNBlock1(payload))$$
$$CNNBlock(x) = LeakyReLU\left(BatchNorm(1DCNN(x))\right) \quad (3)$$

where $PayloadFeature \in \mathbb{R}^{16 \times 21}$, the four parameters (input channels, output channels, kernel size, stride) of 1D-CNN in CNN Block1 and CNN Block2 are (1,1,4,2) and (1,1,3,3) respectively.

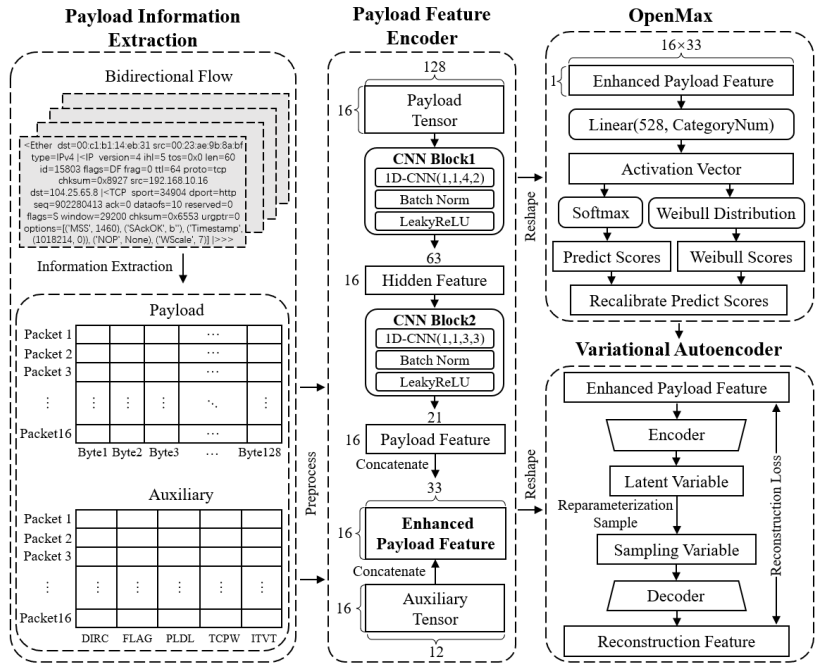

Fig. 1. Overall architecture of VAEMax model. The overall architecture of VAEMax model can be divided into four components (marked by dashed lines), where are: Payload Information Extraction, Payload Feature Encoder, OpenMax, Variational Autoencoder.

Finally, we add auxiliary information to the payload features to enhance their representational capabilities, which can be formalized as follows:

$$EnhancedPayloadFeature = Concatenate(PayloadFeature, Auxiliary) \quad (4)$$

where $EnhancedPayloadFeature \in \mathbb{R}^{16 \times 33}$.

### D. OpenMax

In this subsection, we make some adjustments to the original OpenMax [28] model to adapt to the requirements of this task. We use known flows classification tasks to train the classifier and payload feature encoder. The classifier consists of a linear layer and a softmax layer, and its input is the enhanced payload feature. It should be noted that the dimension of the enhanced payload feature must be reshaped into a 528-dimensional vector before input (16×33=528). The loss function uses cross entropy. After training is completed, we use the activation vector of the classifier (i.e., the output of the linear layer) to fit the Weibull distribution. The fitting process of the Weibull distribution and the recalibration process of prediction scores are shown in Algorithm 1 and Algorithm 2 respectively.

---

**Algorithm 1:** Per category Weibull fit to $\eta$ largest distance to mean activation vector. Returns libMR models $\rho_j$ which includes parameters $\tau_j$ for shifting the data as well as the Weibull shape and scale parameters: $\kappa_j, \lambda_j$.

**Input:** Activation vectors for each category. $ActVector = \{AV_1, \cdots, AV_N\}, AV_j = \{av_1^j, \cdots, av_{N_j}^j\}, av_i^j \in \mathbb{R}^N$, where $ActVector$ is the set of all activation vectors, $AV_j$ is the set of activation vectors of a certain category, $av_i^j$ is the activation vector corresponding to a certain flow, its dimension is the same as the number of categories. $N$ is the number of categories, $N_j$ is the number of flows of category $j$.

**Output:** The mean activation vector $\mu_j$ and libMR model parameters $\rho_j$ for each category.

1: **for** $j = 1$ to $N$ **do**
2:  Compute mean AV, $\mu_j = \text{mean}(av_1^j, \cdots, av_{N_j}^j)$
3:  Compute distance between $\mu_j$ and AV
    $Dist_j = \{\|av_i^j - \mu_j\|\}, i = 1 \dots N_j$
4:  Sort distances from largest to smallest
    $Dist_j = \text{sort}(Dist_j)$
5:  Compute tail length, $\eta_j = \max(0.05 * N_j, 10)$
6:  Fit $\rho_j = (\tau_j, \kappa_j, \lambda_j) = \text{FitHigh}(Dist_j[0:\eta_j], \eta_j)$
7: **end for**

---

Different from the original OpenMax, we modified the calculation method of the tail length, which will be adaptively adjusted according to the amount of data, and should be at least no less than 10. After experiments, it was found that this method is better than using a fixed tail length.

**Algorithm 2:** OpenMax probability estimation with rejection of unknown attack flow.

---

**Require:** Mean AVs $\mu_i$ and libMR models $\rho_i, i = 1 \dots N$.
**Input:** Activation vector $v(x) = \{v_1, \cdots, v_N\}$ and predict score $s(x) = \{s_1, \cdots, s_N\}$ for flow $x$, attenuation rate $\theta$.
**Output:** The recalibrated prediction scores and the flow category.

1: Let $s(x) = \text{argsort}(v(x)) = \{s_1, \cdots, s_N\}$
2: **for** $j = 1$ to $N$ **do**
3:  $i = s_j$
4:  $d_i = \|v(x) - \mu_i\|$
5:  $\omega_i = WScore(d_i, \rho_i)$
6:  $\hat{v}_i = v_i * \left(1 - \omega_i * \frac{N-j-1}{N} * \theta\right)$
7: **end for**
8: $\hat{v}_0 = \sum_{i=1}^{N}(v_i - \hat{v}_i)$
9: Let $y^* = \text{argmax}_i \hat{v}_i, 0 \le i \le N$
10: Reject flow $x$ if $y^* == 0$

---

Different from the original OpenMax, we add attenuation rate $\vartheta$ to the formula on line 6. In the experiment, we found that without this parameter, the model would classify all flows as unknown attacks.

### E. Variational Autoencoder

Since OpenMax can only detect some unknown attacks, the rest are misclassified into other known flows. Therefore, we build its own variational autoencoder for each category of known flows, their model structure is the same, shown as Fig. 2.

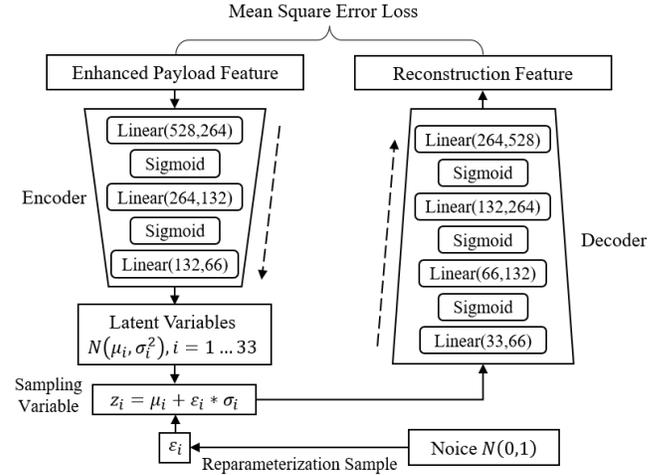

Fig. 2. Model architecture of Variational Autoencoder.

The input to the model is the enhanced payload feature, which is reshape into a 528-dimensional vector. First, use the encoder to extract latent variables. The encoder consists of three linear layers, and the activation function is Sigmoid. The last linear layer outputs a 66-dimensional vector. The first 33 dimensions of this vector constitute the expectation part of the latent variable, and the last 33 dimensions constitute the variance part of the latent variable. Since the vector of the linear output may have negative numbers, and the variance requirement must be a positive number, the last 33 dimensions vector is exponentiated to ensure that it is positive. These 33 pairs of expectations and variances constitute the parameters of the 33 normal distributions. The above process can be formalized as follows:

$$(\mu_i, \sigma_i^2) = (enc_i, e^{enc_{i+33}}), i = 1 \dots 33 \quad (5)$$

where $enc \in \mathbb{R}^{66}$ is the output vector of encoder.

Then use reparameterization trick for data sampling. Since the "sampling" operation is non-differentiable, if the normal distribution $N(\mu_i, \sigma_i^2)$ is sampled directly, the parameters of the encoder cannot be updated through gradient backpropagation. However, the sampling results are differentiable, so we take advantage of the following fact:

- Sampling $z$ from $N(\mu, \sigma^2)$ is equivalent to sampling $\varepsilon$ from $N(0,1)$, and then let $z = \mu + \varepsilon * \sigma$.

Consequently, we will change from sampling from $N(\mu, \sigma^2)$ to sampling from $N(0,1)$, and then obtain the result of sampling from $N(\mu, \sigma^2)$ through reparameterization trick. In this way, the "sampling" operation does not need to participate in gradient descent. Instead, the sampling results participate, making the encoder trainable.

Finally, use the decoder to reconstruct the 33-dimensional sampling vector to 528 dimensions. The encoder consists of four linear layers, and the activation function is Sigmoid. The mean square error is used to calculate the reconstruction loss between the enhanced payload feature and the reconstruction feature, shown as follow:

$$MSELoss = \frac{1}{N}\sum_{i=1}^{N}(epf_i - rf_i)^2, \quad (6)$$

where $epf \in \mathbb{R}^{528}$ is the enhanced payload feature, $rf \in \mathbb{R}^{528}$ is the reconstruction feature, $N = 528$.

For the determination of the reconstruction loss threshold, we adopt the following scheme. For each category, first calculate the reconstruction loss of all data in the training set of this category, then sort them from small to large, and finally select the loss at $\theta * N$ position as the threshold, where $N$ is the data number and $\theta$ is position parameter (default 0.96).

## IV. EXPERIMENT AND ANALYSIS

### A. Datasets

In the experiment, we use CIC-IDS2017 and CSE-CIC-IDS2018 [39] as the benchmark datasets, which are widely used in attack detection. Since the original data is in PCAP format and cannot be used for model training, it needs to be preprocessed. First, we developed a data preprocessing tool, which splits and reassembles packets according to socket information <source IP, source port, destination IP, destination port, protocol>, TCP protocol information <sequence number, acknowledgement number, SYN flag> and packet interval time. Then we generate the label for flows based on attack information <source IP, destination IP, start time, end time>, which can be found on the CIC official website. Due to the excessive number of benign flows and partial attacks, they are randomly sampled. Finally, we randomly selected some attacks as unknown attacks and the rest as known attacks. For the known flows, 80% are randomly selected as the training set and the remaining 20% are used as the testing set, while adding unknown attack flows to the testing set. The specific distribution of the data is shown in Table I and Table II.

TABLE I. CIC-IDS2017 DATASET LABEL DISTRIBUTION

| Category | Train | Test |
|---|---|---|
| Benign | 80,000 | 20,000 |
| FTP Brute Force | 3,177 | 795 |
| DoS Hulk | 8,000 | 2,000 |
| DoS Slowloris | 1,808 | 453 |
| Port Scan | 8,000 | 2,000 |
| DDoS LOIT | 8,000 | 2,000 |
| SSH Brute Force | 0 | 2,508 |
| DoS Slowhttptest | 0 | 1,461 |
| DoS GoldenEye | 0 | 7,567 |
| Botnet | 0 | 736 |
| Web Attack Brute Force | 0 | 72 |
| Web Attack XSS | 0 | 27 |
| Web Attack Sql Injection | 0 | 12 |

TABLE II. CSE-CIC-IDS2018 DATASET LABEL DISTRIBUTION

| Category | Train | Test |
|---|---|---|
| Benign | 48,000 | 12,000 |
| DDoS HOIC | 8,000 | 2,000 |
| DoS Hulk | 8,000 | 2,000 |
| DoS Slowloris | 3,960 | 990 |
| SSH Brute Force | 8,000 | 2,000 |
| Botnet | 0 | 10,000 |
| DDoS LOIC | 0 | 10,000 |
| DoS GoldenEye | 0 | 10,000 |
| Web Attack Brute Force | 0 | 257 |
| Web Attack XSS | 0 | 116 |
| Web Attack Sql Injection | 0 | 44 |

### B. Evaluation Methodology

In order to increase the diversity of evaluation indicators, we evaluate the model performance from two dimensions: binary classification (benign, attack) and multi-classification (benign, known attack 1, …, known attack N, unknown attack). Forbinary classification, use accuracy (ACC), F1-Score (F1), false positive rate (FPR) for model evaluation, shown as (8) to (10). For multi-classification, use the confusion matrix, unknown attack recall rate ($R_{unk}$), weighted precision ($P_{wht}$), weighted recall ($R_{wht}$), and weighted F1-Score ($F1_{wht}$), shown as (11) to (14).

$$ACC = \frac{TP + TN}{TP + FP + FN + TN} \quad (8)$$

$$F1 = \frac{2 * Precison * Recall}{Precison + Recall} \quad (9)$$

$$FPR = \frac{FP_{ben}}{TP_{ben} + FP_{ben}} \quad (10)$$

$$R_{unk} = \frac{TP_{unk}}{TP_{unk} + FN_{unk}} \quad (11)$$

TABLE III. COMPARISON WITH BASELINE

| | CIC-IDS2017 | | | | | | | CSE-CIC-IDS2018 | | | | | | |
|---|---|---|---|---|---|---|---|---|---|---|---|---|---|---|
| | Binary | | | Multi | | | | Binary | | | Multi | | | |
| | $ACC$ | $F1$ | $FPR$ | $R_{unk}$ | $P_{wht}$ | $R_{wht}$ | $F1_{wht}$ | $ACC$ | $F1$ | $FPR$ | $R_{unk}$ | $P_{wht}$ | $R_{wht}$ | $F1_{wht}$ |
| CAVE-EVT[2] | **0.8362** | **0.8572** | 0.0260 | 0.0740 | 0.7103 | 0.6965 | 0.6189 | 0.7785 | 0.6805 | 0.0287 | 0.0081 | 0.4732 | 0.3816 | 0.2817 |
| MetaMax[38] | 0.8104 | 0.8418 | **0.0001** | 0.3065 | 0.6340 | 0.6487 | 0.5924 | 0.7767 | **0.6847** | 0.0018 | 0.6379 | 0.7228 | 0.6816 | 0.6569 |
| VAEMax | 0.8210 | 0.8462 | 0.0245 | **0.4223** | **0.8311** | **0.8052** | **0.7826** | 0.7813 | 0.6800 | 0.0429 | **0.6612** | **0.8605** | **0.7745** | **0.7853** |

TABLE IV. ABLATION EXPERIMENT RESULT

| | CIC-IDS2017 | | | | | | | CSE-CIC-IDS2018 | | | | | | |
|---|---|---|---|---|---|---|---|---|---|---|---|---|---|---|
| | Binary | | | Multi | | | | Binary | | | Multi | | | |
| | $ACC$ | $F1$ | $FPR$ | $R_{unk}$ | $P_{wht}$ | $R_{wht}$ | $F1_{wht}$ | $ACC$ | $F1$ | $FPR$ | $R_{unk}$ | $P_{wht}$ | $R_{wht}$ | $F1_{wht}$ |
| VAEMax | **0.8210** | **0.8462** | 0.0245 | **0.4223** | **0.8311** | **0.8052** | **0.7826** | **0.7813** | 0.6800 | 0.0429 | **0.6612** | 0.8605 | 0.7745 | **0.7853** |
| w/o VAE | 0.7804 | 0.8212 | **0.0007** | 0.2541 | 0.8291 | 0.7647 | 0.7162 | 0.7802 | **0.6879** | 0.0024 | 0.3345 | 0.8495 | 0.5894 | 0.5883 |
| w/o OpenMax | 0.7804 | 0.8177 | 0.0242 | 0.1820 | 0.7690 | 0.7314 | 0.6718 | 0.7813 | 0.6802 | 0.0425 | 0.6611 | **0.8607** | **0.7746** | 0.7853 |

$$P_{wht} = \frac{\sum_{i=0}^{N}(w_i * Precision_i)}{N} \quad (12)$$

$$R_{wht} = \frac{\sum_{i=0}^{N}(w_i * Recall_i)}{N} \quad (13)$$

$$F1_{wht} = \frac{2 * P_{wht} * R_{wht}}{P_{wht} + R_{wht}} \quad (14)$$

## C. Result and Analysis

We use the training data to fine-tune the hyper parameters, including neural network structure selection and OpenMax additional parametrization, and the test data is excluded. Specifically, after each epoch of training, we use all training data to verify the model effect, and fine-tune the hyper parameters by observing the convergence speed and fluctuation degree of the evaluation indicators.

We present the unknown attack detection results in the form of normalized confusion matrix, as shown in Fig. 3 and Fig. 4. The rows of the confusion matrix indicate the predicted labels, and the columns indicate the ground truth labels. The evaluation indicator results are shown in Table III. From the above chart we can find, VAEMax achieves more than 95% classification accuracy for known attacks on two dataset. Meanwhile, for the unknown attacks, VAEMax achieves approximately 42% and 66% recall rate on two dataset respectively. Finally, we also keep the false positive rate of benign flows below 5%. In the performance comparison, we implement the CAVE-EVT and MetaMax as baselines. Table III presents the results of the comparison experiments and the best performance is shown in bold. It can be observed that our proposed model achieves similar performance to baselines in terms of binary classification indicators. However, for multi-classification, our proposed model achieves better performance on all indicators, especially on the recall rate of unknown attack. This shows that the ability of baseline models to identify unknown attacks as attacks is almost the same as our proposed model, but its ability to further classify them as unknown attacks is not as good as our proposed model.

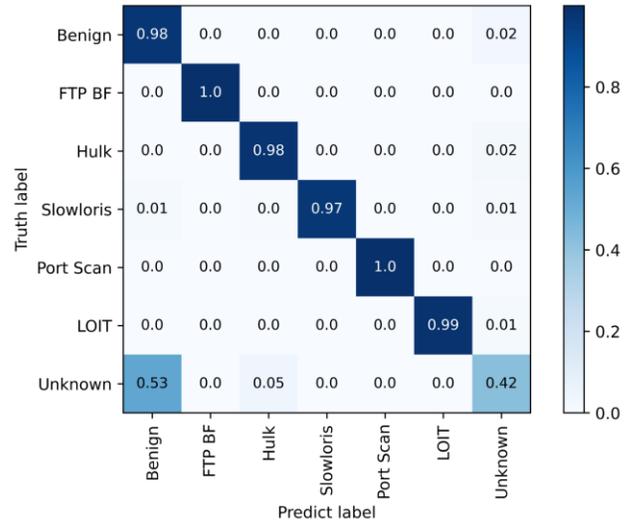

Fig. 3. Normalized confusion matrix of CIC-IDS2017.

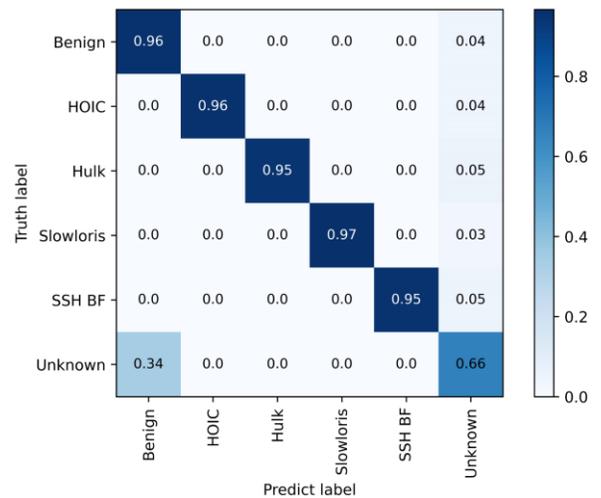

Fig. 4. Normalized confusion matrix of CIC-IDS2018.

TABLE V. STABILITY EXPERIMENT RESULT

| Train Data Percentage | CIC-IDS2017 | | | | | | | CSE-CIC-IDS2018 | | | | | | |
|---|---|---|---|---|---|---|---|---|---|---|---|---|---|---|
| | Binary | | | Multi | | | | Binary | | | Multi | | | |
| | $ACC$ | $F1$ | $FPR$ | $R_{unk}$ | $P_{wht}$ | $R_{wht}$ | $F1_{wht}$ | $ACC$ | $F1$ | $FPR$ | $R_{unk}$ | $P_{wht}$ | $R_{wht}$ | $F1_{wht}$ |
| 80% | **0.8210** | **0.8462** | **0.0245** | **0.4223** | **0.8311** | **0.8052** | **0.7826** | **0.7813** | **0.6800** | **0.0429** | **0.6612** | **0.8605** | **0.7745** | **0.7853** |
| 10% | 0.7591 | 0.8024 | 0.0304 | 0.2885 | 0.7546 | 0.7423 | 0.7002 | 0.7711 | 0.6650 | 0.0647 | 0.6557 | 0.8266 | 0.7474 | 0.7562 |
| 5% | 0.7224 | 0.7796 | 0.0272 | 0.1894 | 0.6325 | 0.6629 | 0.6053 | 0.6653 | 0.5574 | 0.1325 | 0.5105 | 0.7129 | 0.6169 | 0.6170 |

## D. Ablation Experiments

Apart from the performance experiments, we also remove variational autoencoder (named w/o VAE), OpenMax (named w/o OpenMax) from VAEMax to verify the effectiveness of each module respectively. The results are shown in Table IV, where the best performance is marked in bold. It can be found that after removing any module, the recall rate of unknown attacks and the false positive rate in both data sets decrease. In addition, in the CICIDS2017 data set, the complete VAEMax model achieves the best results in all indicators except FPR, and has a large lead. In the CSE-CIC-IDS2018 data set, after removing VAE, the model performance dropped significantly and after removing OpenMax, the model performance change slightly. All above facts indicate that these two modules improve the approach performance and are thus effective.

## E. Stability Experiments

In order to verify the stability of the proposed algorithm, we lower the training data from the original 80% to 20%, 10%, and 5% respectively, and the testing data remain unchanged. The experimental results are shown in Table V. As the training data gradually decreases, the performance of the model deteriorates. But even with only 5% of the training data, the binary classifications accuracy of the model are still 72.24% and 66.53%, which is about 12% and 15% lower than the results of 80% of the training data. Thus, the experimental result indicates that the proposed model has a certain degree of stability.

## V. CONCLUSION

In this paper, we discussed the open-set intrusion detection problem which aimed at identifying more unknown attacks while maintaining the classification accuracy of known flows. By employing the OpenMax and VAE, we proposed VAEMax, a dual detection model to solve this problem. First, OpenMax is used for preliminary classification, and then VAE is used for further detection. Experimental results demonstrated the effectiveness and feasibility of the techniques applied in our VAEMax approach. Currently, we can only classify different types of unknown attacks into the same category. In the future research, we will conduct in-depth research on how to solve multi-classification problems with unknown classes.

ACKNOWLEDGMENT

This work was supported by the National Key R&D Program of China [2022YFB3104300, 2022YFB3104301].